\documentclass[aps,prb,reprint,superscriptaddress]{revtex4-2}

\usepackage{graphicx}
\usepackage{dcolumn}
\usepackage{bm}

\usepackage{appendix}
\usepackage{amsmath,amssymb,amsthm}
\usepackage{lipsum}
\usepackage{mathtools, nccmath,bm}
\usepackage{comment}
\usepackage[utf8]{inputenc}
\usepackage{pgfplots}
\pgfplotsset{compat=1.18}
\usepackage{tikz}
\usepackage{float}
\usepgfplotslibrary{groupplots}
\usepackage{caption} 
\usepackage{booktabs}

\begin{document}


\title{A Generalized Approach to the Relaxation Time of Interacting Magnetic Nanoparticles: From Superparamagnetism to Glassy dynamics}

\author{Jean Claudio Cardoso Cerbino}
\author{Diego Muraca}
\email{dmuraca@unicamp.br}
\homepage{https://sites.ifi.unicamp.br/dmuraca/en} 
\altaffiliation{Also corresponding researcher of the National Scientific and Technical Research Council (CONICET), Argentina.}
\affiliation{“Gleb Wataghin” Institute of Physics, State University of Campinas, SP, Brazil.}
\begin{abstract}
A novel theoretical expression for the relaxation time of magnetic nanoparticles with dipolar interactions is derived from Kramers’ theory, extending the Boltzmann–Gibbs framework to incorporate Tsallis statistics. The model provides a unified description of magnetic relaxation from weakly to strongly interacting regimes. It accounts for both the decrease and the increase of the relaxation time with increasing dipolar coupling, addressing a long-standing problem in nanoparticle magnetism that cannot be consistently described by classical phenomenological models. This result also offers an innovative interpretation of the cut-off condition inherent to the Tsallis distribution in terms of a cut-off temperature, $T_{cut-off}$, which naturally characterizes the onset of glassy freezing dynamics and provides an alternative interpretation of experimental relaxation data within a non-extensive statistical framework.
\end{abstract}

\maketitle

\section{Article Text}
\subsection{Introduction}

\definecolor{accessibleBlue}{RGB}{0, 114, 178}
\definecolor{accessibleOrange}{RGB}{230, 159, 0}
\definecolor{accessibleTeal}{RGB}{0, 158, 115}
\definecolor{accessibleRed}{RGB}{213, 94, 0}
\definecolor{accessiblePurple}{RGB}{148,103,189}

The study of nanomagnetism in single domain magnetic-nanoparticles (MNPs) began with Néel's, further advanced by Brown and other key contributors \cite{Neel1949,Neel1950,Brown1979, Brown1963, AHARONI19923, Stoner1948, Beam, Dormann1997, Coffey1998}. The magnetic moment vector of a MNP is determined by its composition, structure, and temperature, while its orientation is affected by crystalline anisotropy, magnetostatic, and external fields, including interactions with other MNPs. Néel \cite{neel1949influence} predicted that when thermal agitation $k_B T$) equals the energy required to reverse an MNP's magnetic moment ($\bm{\mu}$), its magnetization changes continuously. Delays in magnetization due to external magnetic field changes ($\bm{H_{ext}}$) are termed the "after effect" or "Néel relaxation time" for MNPs.   When the energy barrier $\Delta V(\theta, \phi) = K(\theta, \phi)\nu$ ( with $K(\theta, \phi)$ the anisotropy energy density, $(\theta, \phi)$ the anisotropy directions, and $\nu$ the MNP volume) is smaller than $k_B   T$, the magnetic moments $\boldsymbol{\mu}$ of the MNPs are governed by the reversal relaxation time $\tau$~\cite{Beam}. Néel derived $\tau$, the decay factor for the remanent magnetization 
$M_r(t)=M_s\exp\left(-\frac{t}{\tau}\right)$, where $M_s$ is the saturation magnetization. For single-domain particles, $\tau$ follows the Arrhenius–Néel law $\tau=\tau_0\exp\left(\frac{K\nu}{k_B T}\right)$ with $\tau_0$ being the material-dependent attempt time (characteristic microscopic timescale). 
Conversely, Kramers’ theory \cite{Kramers1940} describes the thermal activation of a particle, such as an MNP in a two-level potential well, by thermal fluctuations from the environment. The particle reaches equilibrium under the Boltzmann–Gibbs (BG) distribution, but may occasionally acquire sufficient energy to cross the potential barrier, temporarily altering the local BG distribution; hence, the core assumption of Kramers’ theory is that the system follows BG statistics.  In the framework of Kramers' theory, the escape rate ($\Gamma$) is expressed as $\Gamma = \Lambda e^{-\frac{\Delta V} {k_B   T}}$,  where $\Delta V$  is the energy barrier. A primary emphasis within this theoretical context is the determination of the prefactor $\Lambda$, which is essential for a comprehensive understanding of the relaxation phenomenon. Furthermore, Kramers' theory facilitates the comprehensive derivation of the phenomenological Néel approximation. Key reviews of Kramers' theory are available in the works of P. Hänggi et al., and by W. T. Coffey and Yuri P. Kalamykov \cite{RevModPhys.62.251, Thelangevin}. Brown \textit{et al.} formulated a dynamic theory to examine both statistical equilibrium and transitional states concerning Brownian motion \cite{Brown1963, 50yearsbrown, 1353448}. This involves the Langevin equation, which includes a random force from thermal fluctuations and Landau–Gilbert–Brown (LGB) equation. In particular, the Fokker–Planck equation for the LGB model was derived under the assumption of a Boltzmann-Gibbs (BG) energy distribution, expressed as $ p(E) \sim e^{-\beta E} $, for particles confined within potential wells \cite{Brown1963}, where $\beta$ is the inverse thermal energy (usually defined as the inverse temperature parameter). A unit sphere model was employed to represent the orientation of magnetic moments, whereby points on the surface of the sphere correspond to the position of unit vectors. Consequently, the LGB equation \cite{1353448} was reformulated to incorporate a stochastic field $\bm{h}(t)$, which arises due to thermal fluctuations \cite{50yearsbrown}.

Dipolar interactions are inherent to any real MNP systems and are critical in determining the magnetic properties of these systems under the influence of an external magnetic field.\cite{Morup1994,Branquinho2013,Landi2014,Moscoso-Londoño2017105, Serantes201828, Goya2004673}. Furthermore, empirical evidence suggests the occurrence of spin glass (SG) or SG-like states, which are marked by a shift from Néel–Brown-type superparamagnetic relaxation to a power-law behavior characteristic of SG systems \cite{Morup1995SpinGlassMaghemite,Luo1991DipoleFrozenFerrofluid,Dormann1998,Mamiya1998Blocking,Djurberg1997}. These states, which appear to result from interactions between MNPs, remain challenging to explain with current theoretical models. Stoner and Wohlfarth suggested the use of the Vogel-Fulcher (VF) law in magnetic interacting systems in the context of Spin-Glass \cite{ShtrikmanWohlfarth1981} , which later became known as the $T^*$ model \cite{Allia2001}. In the $T^{*}$ context, $ k_B   T$ in the Néel relaxation time equation is replaced with $k_B   (T - T^{*})$, incorporating dipolar interactions and the concept of an apparent magnetic moment at the freezing temperature $T^{*}$ proportional to a mean square dipolar field $<B_i^2>$. Dormann \textit{et al.}. \cite{Dormann1988} introduced a model (DBF) where the dipolar interaction among neighboring particles modulates the energy barrier, yielding \(\Delta E = E_a + E_i\), where \(E_i\) accounts for interparticle coupling. Subsequently, Mørup and Tronc (MT) suggested that interactions induce a random dipolar field, resulting in a reduction of the effective energy barrier \cite{Morup1994}. Both the DBF and VF  models forecast an increase in the energy barrier as the interaction strength increases, while the MT model anticipates a decrease in the energy barrier under weak interaction conditions. These models are obtained from the traditional Néel-Brown relaxation time expression derived from Kramers' theory. They integrate interactions either through the modulation of energy barriers $\Delta E$, resulting in an increase (as in the DBF model) or a decrease (as in the MT model), or through the modification of 
the thermal fluctuation parameter $ k_B  T $, as executed in the VF law or $T^{*}$ model.

When interparticle interactions are negligible, the relaxation of MNPs  follows the standard Néel–Brown law as derived from escape‐rate theory: a single energy barrier \(\Delta E\) (or a narrow distribution of barriers) leads to a narrow distribution of relaxation times and an Arrhenius‐type plot that yields physically reasonable values for the anisotropy constant and the prefactor \(\tau_0\). Additionally, the introduction of corrections via the VF, MT, and DBF methodologies could potentially bring the phenomenological results into closer agreement with the experimental observables in weak interaction systems. The influence of dipolar interactions on the magnetism of MNPs has been extensively investigated using models such as Landau–Lifshitz–Gilbert simulations and mean-field theory, aiming to provide key insights into dipolar coupling and its impact on relaxation dynamics \cite{Chalifour2021,Ilg2022,Gallina2023,Anand2021,Porro2019,MunozMenendez2020,GarciaAcevedo2023,PhysRevB.101.224429,Camley2024}. Stochastic and Monte Carlo simulations have further been employed to elucidate the transition from Arrhenius-type relaxation to superspin-glass behavior \cite{Shand2024,Orendac2023,Anand2021,Barrera2021}. In addition, mesoscopic and statistical approaches have been proposed to account for local variations in effective parameters—such as inverse temperature—arising from interparticle interactions and spatial heterogeneity \cite{Chalifour2021,Sadat2023,Barrera2021,Peddis2021}. 

While the mentioned theoretical/phenomenological approaches account for interactions in different ways, they all remain constrained by the Néel–Brown framework, which assumes a BG distribution for the system’s energy, even in the presence of significant interparticle interactions. However, as correlations among magnetic moments increase, the BG statistical framework may fail to accurately characterize the system's energy distribution. This inadequacy stems from fluctuations in the inverse temperature parameter $ \beta $. Locally, within spatial "cells," $ \beta $ can be seen as nearly constant, making the energy distribution in each cell follow the usual Boltzmann factor $ e^{-\beta E} $, with $ E $ representing the effective energy. Addressing fluctuations requires averaging over spatial-temporal variations of $ \beta $, producing two statistical contributions: one from the distribution of $ \beta $ and another from the local Boltzmann factor $ e^{-\beta E} $, as described in \textit{superstatistics} by Beck and Cohen \cite{superstatistics}.
For moderate to strong interactions, the inadequacy of the Néel–Brown model becomes evident, as it fails to generate valid physical parameters through straightforward barrier adjustments, as initially observed by Dormann \cite{Dormann1999}. In this regime, nanoparticles can be characterized as \textit{superspins} within a strongly correlated ensemble, manifesting glass-like characteristics, magnetic aging, memory effects, and critical slowing down, which closely resemble the dynamics observed in spin glasses \cite{PhysRevLett.91.167206, PhysRevB.102.174449, PhysRevB.71.104405}. Hence, one often applies a  spin‐glass critical dynamics law, valid only in the vicinity of the glass transition, i.e $T\rightarrow T_g$, $   \tau = \tau^{*} \, \left(\dfrac{T_g}{T-T_g}\right)^{z\nu}\,,
$ 
where $z\nu$ is a critical exponent, and $\tau^{*}$ is the flip time of a single spin, whereas for MNPs it corresponds to the flip time of the magnetic moments. This letter introduces a generalized theory based on Brown's formalism for interacting MNP systems, extending the Boltzmann–Gibbs framework to the \textit{Tsallis statistics}, as first proposed in 1988 \cite{tsallis1988}. Within this framework, entropy assumes a non-extensive nature, resulting in the emergence of \textit{non-extensive statistical mechanics}. This leads to a generalized energy distribution, referred to as the \textit{q}-distribution: $p(E) \propto e_q^{-\beta'_q E} = \left[1 - (1 - q)\beta'_q E \right]^{\frac{1}{1 - q}} $.  Here, $ q $ serves as the entropic index, which quantifies the degree of non-extensivity and links microscopic fluctuations that are difficult to account for to the emergent macroscopic statistics. Specifically, $ q = 1 $ retrieves the conventional Boltzmann-Gibbs statistics, whereas $ q \neq 1 $ accounts for deviations engendered by correlations, memory effects, or long-range interactions.  This formalism has proven successful in describing various complex systems \cite{WilkWlodarczyk2000, Beck2001, Sakaguchi2001, Beck2003Turbulence,Pickup2009,Tsallis2003CosmicRays}.

\subsection{Theory and Results}

To account for interparticle interactions, we consider Brown's formalism of the unitary sphere model, where an MNP magnetic moment \( \bm{u} \) is represented by a point with angular coordinates \( (\theta, \phi) \). This unit vector's behavior follows the stochastic (LGB) equation:
\begin{equation}\label{Landau}
    \dot{ \bm{u}}(t) = \gamma \left[ \left( \mathbf{H}(t) + \mathbf{h} - \frac{\alpha}{\gamma} \dot{\bm{u}}(t) \right) \times \bm{u}(t) \right],
\end{equation}
where \( \alpha \) is the damping constant, \( \gamma \) is the gyromagnetic ratio, and $\bf{H}(t)$ and $\bf{h}(t)$ are the applied and random magnetic fields, respectively. Assuming the continuity equation for the probability density function \( W(\theta, \phi, t) \) on the unit sphere, a corresponding Fokker–Planck equation can be derived. Contrary to the traditional methodology, we do \textit{not} postulate an isotropic diffusion coefficient in Fick's law $J_{diff}=-k' \nabla W$ due to the induced fluctuations in interactions, as previously elaborated.

Under conditions of uniaxial symmetry, the Fokker–Planck equation takes on the following form:
\begin{equation}\label{generalfokkeruniaxial}
\begin{split}
\frac{\partial W}{\partial t} ={} & \frac{h'}{\sin \theta} \frac{\partial}{\partial \theta} \left( \sin \theta \frac{d V(\theta)}{d\theta} W \right) \\
&  \hspace{1cm} +\frac{1}{\sin \theta} \frac{\partial}{\partial \theta} \left( \sin \theta \, k' \frac{\partial W}{\partial \theta} \right),
\end{split}
\end{equation}
where  $ h'={\lambda}/{(\alpha + \alpha^{-1})\mu_0 M_s} $ reflects the deterministic drift component, with $\mu_0$ being Bohr's magneton and $M_s$ the saturation magnetization, $k'$ is the probability of anisotropic diffusion resulting from spatio-temporal fluctuations in the system, and $V(\theta)$ is the potential energy, uniaxial in our case . By considering the Tsallis  energy distribution (see \ref{End.Matter}) as the stationary solution of Eq.  ~\ref{generalfokkeruniaxial} given by $W_0=e_q^{-\beta'_q\left(V(\theta)-V(\theta_0)\right)}/Z$, where $V(\theta_0)$ is the global minimum of potential energy, $Z$ represents the partition function, and $\beta'_q$ denotes the Lagrange multiplier of the internal energy constraint,  the diffusion coefficient $k'$ can be ascertained, leading to:

\begin{equation}
    k'=h'\dfrac{Z^{1-q}}{\beta'_q } W_0^{1 -q}.
\end{equation}

 From the Tsallis–Mendes–Plastino formulation~\cite{Tsallis1998ConstraintsNonextensive}, the $\beta'_q$ parameter respects the following self-consistent relation: \begin{equation}\label{beta}
    \beta'_q = \dfrac{\chi_q^{-1}}{k_B  \left( T + \left(1 - q\right) \dfrac{U_q}{\chi_q k_B } \right)}= \dfrac{\chi_q^{-1}}{k_B  \left( T \pm T^*\right)},
\end{equation}
where + or $-$ signifies $q<1$ or $q>1$, respectively; $\chi_q$ represents a normalization constant defined as $\chi_q = 2\pi \int_{0}^{\pi} W_0^{q}\sin(\theta)\,d\theta$; and $U_q$ signifies the $q$-expectation value of the internal energy defined as:

\begin{equation}\nonumber
    U_q = \dfrac{2\pi \int_{0}^{\pi} W_0^q V(\theta) \sin(\theta)\,d\theta}{\chi_q}, 
\end{equation} and
where, for convenience, the temperature $T^*$ was defined as:
\begin{equation}\label{Tstar}
T^{*} = \frac{|1 - q|\,U_q}{k_B \,\chi_q}.
\end{equation}

The equation ~\ref{generalfokkeruniaxial} can be reformulated appropriately to determine the Fokker-Planck operator $L_{fp}$ and address the mean first passage time problem for the interval $(\theta_i,\theta_c)$, thereby yielding the reversal time for $q<2$:

\begin{equation}\label{relaxtheta}
\begin{split}
\left<T_1\right>(\theta_i) = 2\tau_N \int_{\theta_i}^{\theta_C} 
\dfrac{\left(1 - (1 - q)\beta'_q V(\theta)\right)^{\frac{2 - q}{1 - q}}}
{\sin(\theta)}\,d\theta \\
\times \int_{\theta_i}^{\theta}
\left(1 - (1 - q)\beta'_q V(\theta')\right)^{\frac{1}{1 - q}} \sin(\theta')\,d\theta'
\end{split}
\end{equation}
where $\tau_N=\dfrac{\beta'_q}{2h'}$ is a characteristic time constant. In the high-barrier limit, a saddle point approximation can be made for the potential $V(\theta)=K\nu \sin^2(\theta)$,  where $K$ is the effective anisotropy constant and $\nu$ is the particle's volume. Because most of the contributions of the integrals in Eq~\ref{relaxtheta} come from the points near ($\theta_i,\theta_C$), when $\theta_i$ is one of the wells, i.e., $V(\theta_i)=0$, and $\theta_C$ is the maximum of the potential,  the reversal time asymptotically approaches: 

\begin{equation}\label{q-Assimptotic}
    \tau_i = \tau_0\left(1-(1-q)\beta'_qV(\theta_C) \right)^{\frac{q-3}{2(1-q)}},
\end{equation}
where \( \tau_0 \) is defined as:
\begin{equation}
    \tau_0 = \frac{\tau_N \sqrt{2} C_1}{\sin(\theta_C)(\beta'_{q})^{3/2} V''(\theta_i)(2 - q)} \cdot \frac{1}{\sqrt{ |V''(\theta_C)|}},
\end{equation}
with $V''(\theta)$ being $d^2V(\theta)/d\theta^2$ and the constant \( C_1 \) is given by:

\begin{equation}\label{mres}
C_1 =
\begin{cases} 
\frac{\sqrt{\pi} \, \Gamma\left(\frac{1}{2} + \frac{1}{1 - q}\right)}{\sqrt{1 - q} \, \Gamma\left(\frac{q - 2}{q - 1}\right)}, & \text{if } q < 1, \\[10pt]
\sqrt{\pi}, & \text{if } q = 1, \\[10pt]
\frac{\sqrt{\pi} \, \Gamma\left(\frac{1}{q - 1}\right)}{\sqrt{q - 1} \, \Gamma\left(\frac{1}{2} + \frac{1}{q - 1}\right)}, & \text{if } q > 1.
\end{cases}
\end{equation}
For $q=1$, Eq  .~\eqref{q-Assimptotic} matches Brown's classical asymptotic formula. The same result given by this equation can be obtained by accounting for deviations from normal diffusion due to interactions through a generalized form of Fick’s law $J = -k' \nabla W^\nu$  extensively studied in nonextensive statistical mechanics, resulting in Tsallis energy distributions as stationary solutions in porous media diffusion processes \cite{Plastino}. In accordance with these diffusion processes, $q > 1$ denotes superdiffusive behavior, whereas $q < 1$ pertains to subdiffusion \cite{Borland1998,Tsallis1996}. Normal diffusion is recovered in the limit $q = 1$. By comparing with diffusion processes, our results, as presented in Eq.~\ref{q-Assimptotic}, exhibit an analogy with the generalized theory of the Arrhenius law specific to porous media, as proposed by Lenzi \textit{et al.}~\cite{Lenzi2001}.

In particular, the Tsallis formalism for $q<1$ imposes an abrupt cut-off to ensure the positivity of the distribution, i.e. $(1 - (1 - q)\beta'_q V(\theta_C)) > 0$. From this condition, a cut-off temperature $T_\mathrm{cut\mbox{-}off}$ can be defined, below which the probability distribution becomes identically zero:
\begin{equation}\label{T-cutoff}
T_{\mathrm{cut\mbox{-}off}}
= \frac{|1 - q|}{k_B \chi_q}\bigl(\Delta V - U_q\bigr).
\end{equation}
where $\Delta V=V(\theta_C)$ is the energy barrier. Although  $T_{\mathrm{cut\mbox{-}off}}$ was derived when $q<1$, it can be employed to reformulate Eq.~\eqref{mres}, resulting in three distinct regimes:
\begin{equation}\label{model}
\tau = 
\begin{cases}
\tau_0\displaystyle\left(\frac{T + T^*}{T - T_{\mathrm{cut\mbox{-}off}}}\right)^{\frac{3 - q}{2(1 - q)}},
& q < 1,\\[1ex]
\tau_0 \exp\!\bigl(K\nu/(k_B  T)\bigr),
& q = 1,\\[1ex]
\tau_0\displaystyle\left(\frac{T - T^*}{T + T_{\mathrm{cut\mbox{-}off}}}\right)^{\frac{3 - q}{2(1 - q)}},
& q > 1,
\end{cases}
\end{equation}

 For $q < 1$, the relaxation time $\tau$ diverges at $T = T_{\mathrm{cut\mbox{-}off}}$ due to the abrupt energy cut-off, leading to thermodynamic freezing. For $q = 1$, the standard Boltzmann–Gibbs statistics is recovered, with $\tau$ following a classical Néel-Brown behavior. For $q > 1$, the distribution acquires a long power-law tail instead of a cut-off and remains positive across all accessible energy levels. In this case, no such transition temperature exists, and consequently, no divergence of $\tau$ occurs.

\begin{figure}[H]
  \centering
  \begin{tikzpicture}
    \begin{groupplot}[
      group style={
        group name=two plots,
        group size=1 by 2,
        horizontal sep=2cm,
        vertical sep=1.2cm
      },
      width=0.4\textwidth,
      height=0.4\textwidth,
      ymin=0, ymax=200,
      grid=both,
      grid style={dashed, gray!30},
      tick align=outside,
      scaled x ticks=false,
      xlabel near ticks,
      ylabel near ticks,
      legend cell align={left},
      clip=true,
      clip marker paths=true
    ]

    \nextgroupplot[
      xlabel={$T$ (K)},
      ylabel={$T^*$ (K)},
      xmin=0, xmax=250,
      ymin=0, ymax=80,
      legend style={
        at={(0.03,0.97)},
        anchor=north west,
        draw=none,
        fill=white,
        font=\footnotesize
      }
    ]
      \addplot[mark=*, mark options={fill=accessibleBlue}, solid, line width=1pt, color=accessibleBlue]
        table [x index=1, y index=2] {Temperaturas/Q07_Temperaturas_Tfreezing_Testrela.txt};
      \addlegendentry{$q=0.7$}

      \addplot[mark=square*, mark options={fill=accessibleOrange}, solid, line width=1pt, color=accessibleOrange]
        table [x index=1, y index=2] {Temperaturas/TabelaExtendidaQ08_beta_Tphys_Testrela_Tfreezing.dat};
      \addlegendentry{$q=0.8$}

      \addplot[mark=diamond*, mark options={fill=accessibleTeal}, solid, line width=1pt, color=accessibleTeal]
        table [x index=1, y index=2] {Temperaturas/TabelaExtendidaQ09_beta_Tphys_Testrela_Tfreezing.dat};
      \addlegendentry{$q=0.9$}

      \addplot[mark=triangle*, mark options={fill=accessiblePurple}, solid, line width=1pt, color=accessiblePurple]
        table [x index=1, y index=2] {Temperaturas/TabelaExtendidaQ95_beta_Tphys_Testrela_Tfreezing.dat};
      \addlegendentry{$q=0.95$}

      \addplot[mark=pentagon*, mark options={fill=accessibleRed}, solid, line width=1pt, color=accessibleRed]
        table [x index=1, y index=2] {Temperaturas/TabelaExtendidaQ98_beta_Tphys_Testrela_Tfreezing.dat};
      \addlegendentry{$q=0.98$}


    \nextgroupplot[
      xlabel={$T$ (K)},
      ylabel={$T_{cut-off}$ (K)},
      xmin=0, xmax=250,
      ymin=0, ymax=120,
      yticklabel style={/pgf/number format/fixed},
      legend style={draw=none}
    ]

      \addplot[mark=*, mark options={fill=accessibleBlue}, solid, line width=1pt, color=accessibleBlue]
        table [x index=1, y index=3] {Temperaturas/Q07_Temperaturas_Tfreezing_Testrela.txt};

      \addplot[mark=square*, mark options={fill=accessibleOrange}, solid, line width=1pt, color=accessibleOrange]
        table [x index=1, y index=3] {Temperaturas/TabelaExtendidaQ08_beta_Tphys_Testrela_Tfreezing.dat};

      \addplot[mark=diamond*, mark options={fill=accessibleTeal}, solid, line width=1pt, color=accessibleTeal]
        table [x index=1, y index=3] {Temperaturas/TabelaExtendidaQ09_beta_Tphys_Testrela_Tfreezing.dat};

      \addplot[mark=triangle*, mark options={fill=accessiblePurple}, solid, line width=1pt, color=accessiblePurple]
        table [x index=1, y index=3] {Temperaturas/TabelaExtendidaQ95_beta_Tphys_Testrela_Tfreezing.dat};

      \addplot[mark=pentagon*, mark options={fill=accessibleRed}, solid, line width=1pt, color=accessibleRed]
        table [x index=1, y index=3] {Temperaturas/TabelaExtendidaQ98_beta_Tphys_Testrela_Tfreezing.dat};
         
    \end{groupplot}
  \end{tikzpicture}
  \caption{(first) $T^{*}$  vs temperature  $T$ ; (second) $T_{\mathrm{cut-off}}$ vs temperature for an energy barrier $\Delta V/k_B =555K$ and different values of $q$.}
  \label{fig:Testrela_Tfreezing_sidebyside}
\end{figure}
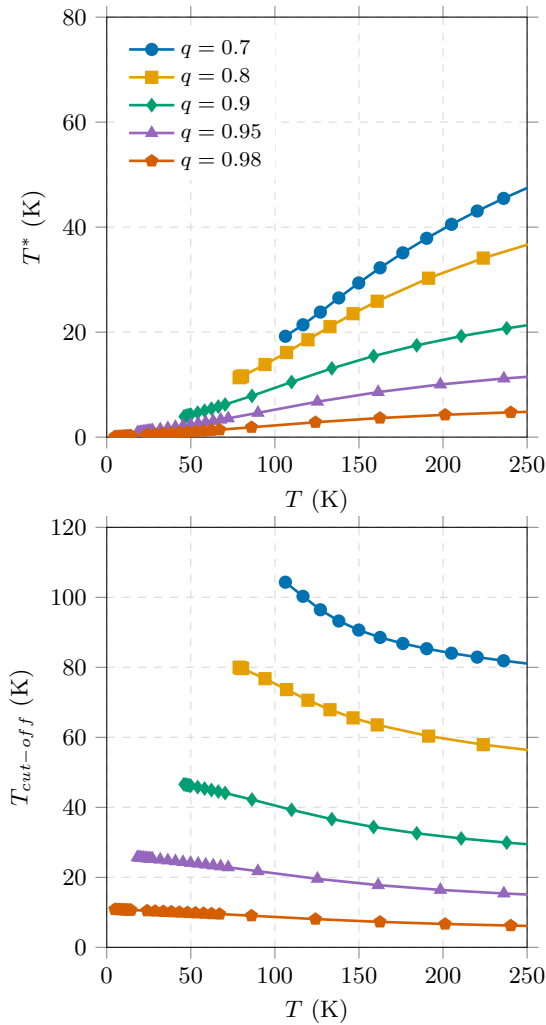
Figure ~\ref{fig:Testrela_Tfreezing_sidebyside} demonstrates the behavior of the temperatures $T^*$ (and $T_{\mathrm{cut-off}}$) as a function of temperature. Figure 
~\ref{fig:Testrela_Tfreezing_sidebyside} (left) shows $T^*$ increase 
(with internal energy ($U_q$)) as temperature increases, while $q$ is constant. It also increases as $q$ decreases at a fixed temperature, showing a direct correlation between $T^*$ and interparticle interactions. Figure ~\ref{fig:Testrela_Tfreezing_sidebyside} (right) shows that $T_{\mathrm{cut-off}}$ decreases as temperature increases. The possible limit  value of $T_{\mathrm{cut-off}}$, linked to the crossover temperature, increases with MNPs interaction strength. 
Hence, the strength of MNPs interaction determines both $T^*$, related to internal energy, and the cut-off temperature $T_{\mathrm{cut-off}}$.

\begin{figure}[H]
  \centering
  \begin{tikzpicture}
    \begin{axis}[
      width=8.5cm,                       
      height=6.375cm,                    
      xlabel={$1/\beta'_q$ (J)},
      ylabel={Temperature (K)},
      grid=both,
      grid style={dashed, gray!30},
      tick align=outside,
      legend cell align={left},
      legend style={
        at={(0.72,0.68)},
        anchor=south east,
        draw=none,                                
        fill=none,
        font=\small,
        /tikz/every even column/.append style={column sep=6pt}
      },
      xmin=1e-21,
      xmax=5e-21,
      ymin=0,
      ymax=250,
    ]

    \addplot[
      color=accessibleBlue,
      mark=*,
      mark options={fill=accessibleBlue, draw=accessibleBlue},
      solid, line width=1.0pt, mark size=2pt
    ]
    table [x expr=1/\thisrowno{0}, y index=1] {Temperaturas/TabelaExtendidaQ07_beta_Tphys_Testrela_Tfreezing.dat};
    \addlegendentry{$T$}

    \addplot[
      color=accessibleBlue,
      mark=square*,
      mark options={fill=white, draw=accessibleBlue},
      dashed, line width=1.0pt, mark size=2pt
    ]
    table [x expr=1/\thisrowno{0}, y index=3] {Temperaturas/TabelaExtendidaQ07_beta_Tphys_Testrela_Tfreezing.dat};
    \addlegendentry{$T_{\mathrm{cut\text{-}off}}$}

    \addplot[
      only marks,
      mark=triangle*,
      mark size=6pt,
      color=black
    ] coordinates {(2.35e-21, 105)};
    \addlegendentry{Crossover temperature}

    \node[anchor=west, font=\scriptsize, fill=white, inner sep=1pt, rounded corners=1pt]
      at (axis cs:2.6e-21, 105) {};

    \end{axis}
  \end{tikzpicture}
  \caption{Temperatures $T$ (solid line) and $T_{\mathrm{cut-off}}$ (dashed line) as a function of $1/\beta'_q$ for q=0.7, indicating the crossover temperature where $T=T_{\mathrm{cut\!-\!off}}$.}
  \label{fig:Q07_T_vs_invbeta}
\end{figure}
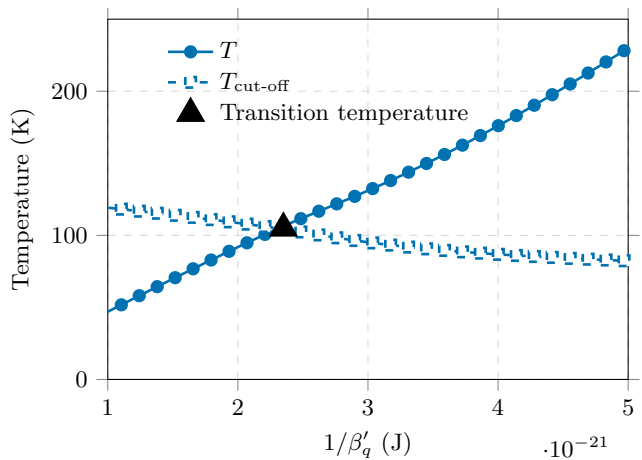

For a more comprehensive understanding of $T_{\mathrm{cut\text{-}off}}$, Fig.~\ref{fig:Q07_T_vs_invbeta} illustrates the relationship between $T_{\mathrm{cut\text{-}off}}$ and the temperature $T$ as functions of $\beta'_q$ (for the arbitrarily selected value $q = 0.7$, chosen to illustrate a representative example of the crossover temperature defined in Eq.~\eqref{beta}). In particular, the normalization constant $\chi_q$ and the $q$-expectation value $U_q$ are $q$-weighted statistical averages; as such, they depend on the occupation probabilities of the microstates and therefore vary with the thermodynamic conditions (temperature), inducing corresponding changes in $\beta'_q$.

When the temperature approaches $T = T_{\mathrm{cut\text{-}off}}$, the system exhibits a crossover from superparamagnetic behavior to a glassy-like dynamical regime, in which the relaxation becomes strongly slowed and the system is effectively frozen, with a loss of ergodic exploration of phase space. In dense assemblies of strongly interacting magnetic nanoparticles, the combination of positional disorder and frustrating dipolar interactions gives rise to a rugged energy landscape that governs slow collective relaxation dynamics. Below the glassy freezing temperature, the system becomes dynamically trapped within local energy minima, exhibiting effectively broken ergodicity. Within this context, Tsallis statistics ($q < 1$) provide an effective description of the restricted phase-space exploration through the emergence of a cut-off condition that limits the maximum accessible energy, representing the barrier that confines the magnetic nanoparticle system within a given energy valley. By assigning vanishing probability to higher-energy states, the $q < 1$ framework captures the system’s confinement and non-ergodic nature.

 Figure ~\ref{fig:numeric_vs_asymptotic} shows the relaxation time as a function of the inverse of temperature, considering a system with an energy barrier of $\Delta V/k_B =555$K, for ultradiluted 4.7 nm $\gamma-$Fe$_2$0$_3$ MNP's, as reported by Dormann \textit{et al.} with different concentrations  \cite{Dormann1998}. The  Eq.~\eqref{relaxtheta} was numerically integrated (symbols)  to represent the expected behavior of the relaxation time, which was subsequently compared to the high barrier approximation using Eq.~\ref{model} (lines). A transition from sub-diffusive to super-diffusive behavior can be observed under the condition $q=1$,  where the model captures classical Néel-Brown non-interacting behavior. Generally, for $q < 1$ (sub-diffusive), reducing $q$ makes the system resemble a strongly interacting nanoparticle ensemble, where the relaxation time increases as $q$ decreases (interactions strengthen) and diverges at a finite critical temperature. The particle size, geometry, and spatial arrangement of MNPs should also affect $q$. Even with similar mean separations, variations in local ordering or magnetic anisotropy can lead to different levels of nonextensivity. A thorough understanding of how these structural factors affect $q$ needs further experimental, simulation, and/or theoretical study.
 
\definecolor{accessibleBlue}{RGB}{0, 114, 178}
\definecolor{accessibleOrange}{RGB}{230, 159, 0}
\definecolor{accessibleTeal}{RGB}{0, 158, 115}
\definecolor{accessibleRed}{RGB}{213, 94, 0}
\definecolor{accessiblePurple}{RGB}{204, 121, 167}
\definecolor{accessibleBlack}{RGB}{0, 0, 0}

\begin{figure}[H] 
\centering
\begin{tikzpicture}\label{fig_model_q}
    \begin{axis}[
        width=8.5cm,
        height=12cm,
        xlabel={$1/T \ (\text{K}^{-1})$},
        ylabel={$\tau/\tau_N $},
        ymode=log, 
        grid=both,
        grid style={dashed, gray!30},
        legend style={
  at={(0.28,0.98)},
  anchor=north east,
  draw=black,
  fill=white,
  font=\footnotesize,
  legend columns=1
},
        xmin=0, 
        xmax=0.025, 
        ymin=1,
        ymax=10^4
    ]

    \addplot[accessibleBlue, solid, line width=1.2pt, forget plot] 
        table {latex_data/Resultq07Asymptotic.dat};
    \addplot[accessibleBlue, only marks, mark=*, mark size=1.5pt] 
        table {latex_data/Resultq07Numeric.dat};
    \addlegendentry{$q=0.7$}

    \addplot[accessibleOrange, solid, line width=1.2pt, forget plot] 
        table {latex_data/Resultq08Asymptotic.dat};
    \addplot[accessibleOrange, only marks, mark=square*, mark size=1.5pt] 
        table {latex_data/Resultq08Numeric.dat};
    \addlegendentry{$q=0.8$}

    \addplot[accessibleTeal, solid, line width=1.2pt, forget plot] 
        table {latex_data/Resultq09Asymptotic.dat};
    \addplot[accessibleTeal, only marks, mark=diamond*, mark size=2pt] 
        table {latex_data/Resultq09Numeric.dat};
    \addlegendentry{$q=0.9$}
    
    \addplot[accessibleBlack, solid, line width=1.2pt, forget plot] 
        table {latex_data/Resultq10Asymptotic.dat};
    \addplot[accessibleBlack, only marks, mark=triangle*, mark size=2pt] 
        table {latex_data/Resultq10Numeric.dat};
    \addlegendentry{$q=1.0$}
    
\node[anchor=west, font=\footnotesize] at (axis cs:0.017, 4e2) {N\'eel--Brown};
    \addplot[accessibleRed, solid, line width=1.2pt, forget plot] 
        table {latex_data/Resultq11Asymptotic.dat};
    \addplot[accessibleRed, only marks, mark=pentagon*, mark size=2pt] 
        table {latex_data/Resultq11Numeric.dat};
    \addlegendentry{$q=1.1$}

    \addplot[accessiblePurple, solid, line width=1.2pt, forget plot] 
        table {latex_data/Resultq13Asymptotic.dat};
    \addplot[accessiblePurple, only marks, mark=+, mark size=2pt, thick] 
        table {latex_data/Resultq13Numeric.dat};
    \addlegendentry{$q=1.3$}
    \end{axis}
\end{tikzpicture}
\caption{A comparative analysis between the numerically calculated relaxation time, Eq~\eqref{relaxtheta} (symbols) and the asymptotic approximation, Eq. \ref{model} (lines), for various values of the parameter $q$. The approximation is more accurate for lower temperatures in the high barrier limit.}
\label{fig:numeric_vs_asymptotic}
\end{figure}
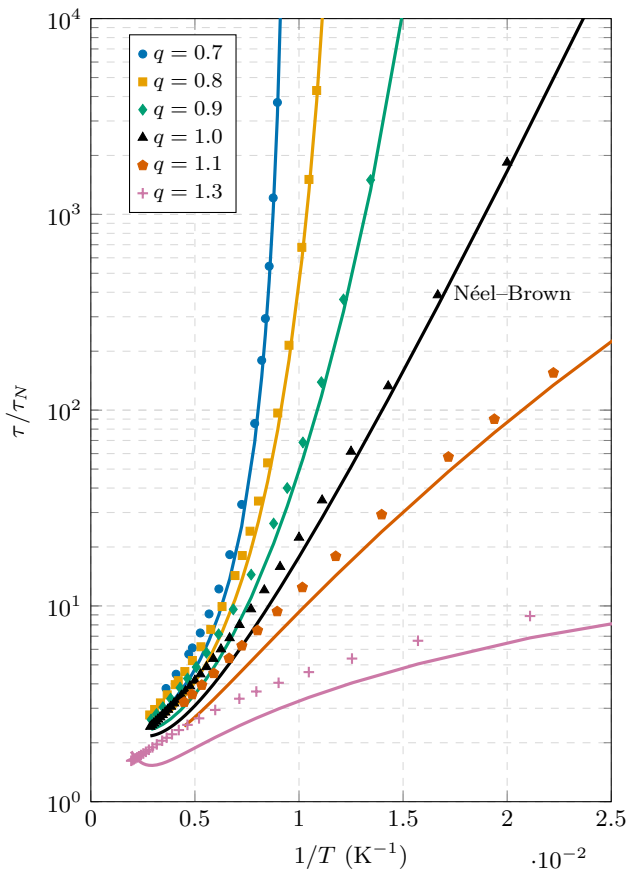
 In the super-diffusive scenario ($q>1$), a reduction in relaxation time is observed as $q$ increases. Reports on super-diffusive regimes in MNPs systems are scarce, indicating that this behavior is not universal but occurs under specific spatial configurations, such as chain-like assemblies, for example \cite{Morup1994,PhysRevB.70.144401,PhysRevLett.88.217205}. These create inhomogeneous dipolar fields, leading to collective magnetization dynamics predicted by nonextensive statistical mechanics. Similar super-diffusion appears in other complex systems only in certain microstructural or dynamic settings. Thus, in the Tsallis framework, $q>1$ behavior signals specific correlated states, not a general system trait.

\subsection{Application of the theory}

 The applicability of the  model was based on classical and well-recognized literature reported on magnetic interaction strength and diverse types of collective behaviors \cite{Dormann1998,Djurberg1997}. Equation \eqref{model} was used to analyze the reported data and to correlate the fitted parameters with theoretical predictions.

\begin{figure}[ht]
  \centering
  \begin{tikzpicture}

    \begin{axis}[
      name=main,
      width=8.5cm,
      height=7.5cm,
      xlabel={$1/T \ (K^{-1})$},
      ylabel={$\log_{10}(\tau)$}, 
      xmin=0, xmax=0.05,
      ymin=-10, ymax=1,
      grid=both,
      grid style={dashed,gray!30},
      tick align=outside,
      tick style={black},
      scaled x ticks=false,
      x tick label style={
        /pgf/number format/fixed,
        /pgf/number format/precision=3
      },
      every axis label/.append style={font=\footnotesize},
      every tick label/.append style={font=\footnotesize},
      legend style={
        at={(0.75, 0.9)}, 
        anchor=north west, 
        font=\footnotesize,
        draw=none,
        fill=none,
      },
      legend cell align={left},
    ]

      \addplot[only marks, mark=*, color=accessibleBlue, mark options={fill=accessibleBlue, scale=0.7}]
        table [x index=0, y index=1] {data/Dataset1Data.dat};
      \addlegendentry{Powder} 
      \addplot[solid, line width=1pt, color=accessibleBlue, forget plot] 
        table [x index=0, y index=1] {data/Dataset1Fit.dat};

      \addplot[only marks, mark=square*, color=accessibleOrange, mark options={fill=accessibleOrange, scale=0.7}]
        table [x index=0, y index=1] {data/Dataset2Data.dat};
      \addlegendentry{Floc} 
      \addplot[solid, line width=1pt, color=accessibleOrange, forget plot] 
        table [x index=0, y index=1] {data/Dataset2Fit.dat};

      \addplot[only marks, mark=diamond*, color=accessibleTeal, mark options={fill=accessibleTeal, scale=0.9}]
        table [x index=0, y index=1] {data/Dataset3Data.dat};
      \addlegendentry{IN} 
      \addplot[solid, line width=1pt, color=accessibleTeal, forget plot] 
        table [x index=0, y index=1] {data/Dataset3Fit.dat};

      \addplot[only marks, mark=triangle*, color=accessibleRed, mark options={fill=accessibleRed, scale=0.9}]
        table [x index=0, y index=1] {data/Dataset4Data.dat};
      \addlegendentry{IF} 
      \addplot[solid, line width=1pt, color=accessibleRed, forget plot] 
        table [x index=0, y index=1] {data/Dataset4Fit.dat};
      

    \end{axis}

    \begin{axis}[
      at={(main.south east)},
      anchor=south east,
      xshift=-7pt, yshift=35pt,
      width=3.8cm,
      height=3.2cm,
      xlabel={$(1/T )$}, 
      ylabel={$\log_{10}(\tau )$}, 
      xmin=0.0, xmax=0.02,
      ymin=-5, ymax=1,
      grid=both,
      grid style={dashed,gray!25},
      tick align=outside,
      tick style={black},
      every axis label/.append style={font=\footnotesize},
      every tick label/.append style={font=\footnotesize},
      axis background/.style={fill=white, draw=black},
      xmajorgrids=false,
      ymajorgrids=false,
    ]

      \addplot[only marks, mark=*, color=accessibleBlue,
        mark options={fill=accessibleBlue, scale=0.6}]
        table [x index=0,y index=1] {data/SamplePSGData.dat};

      \addplot[solid, line width=0.8pt, color=black]
        table [x index=0,y index=1] {data/SamplePSGFit.dat};

    \end{axis}

  \end{tikzpicture}

  \caption{Relaxation time $\tau$ as a function of inverse temperature $1/T$ for samples with increasing concentrations IF, IN, Floc for the conglomerated sample and for a Powder sample. The lines indicate the fit from Eq~\eqref{model}. The relaxation time increases with interaction strength, indicating slower dynamics in more strongly interacting assemblies. An inset displays a fit to the spin-glass equation for the Powder sample yielding $\tau^{*}=10^{-10\pm2}$ s, $z\nu=7\pm3$ and $T_{g} = 130\pm 4\,$K. Data reproduced from Fig.~1 of Dormann \emph{et~al.}~\cite{Dormann1998}.}
  \label{fig:relaxation-fits-four-samples}
\end{figure}
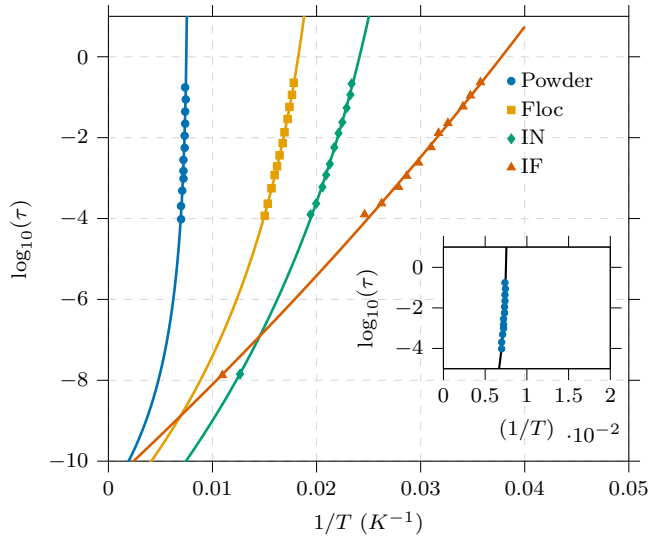

\begin{table}[h!]
\centering
\caption{Parameters obtained from fitting the Eq~\eqref{model}}
\label{tab:fit_parameters}
\begin{tabular}{|l|c|c|c|c|}
\hline
\textbf{Sample} & \textbf{$\log_{10}(\tau_0)$} & \textbf{$T_{cut-off}$ [K]} & \textbf{$T^{*}$[K]} & \textbf{$q$} \\
\hline

Powder & -11 $\pm$ 3 & 130 $\pm$ 4 & 20 & 0.8 $\pm$ 0.1 \\
\hline
Floc   & -11.3 $\pm$ 0.5 & 45 $\pm$ 2 & 2.8 & 0.93 $\pm$ 0.01 \\
\hline
IN     & -12.4 $\pm$ 0.5 & 30.1 $\pm$ 0.9 & 1.0 & 0.954 $\pm$ 0.003 \\
\hline
IF     & -10.5 $\pm$ 0.2 & 7 $\pm$ 2 & 0.1 & 0.986 $\pm$ 0.005 \\
\hline
\bottomrule
\end{tabular}
\end{table}

Figure~\ref{fig:relaxation-fits-four-samples} presents the relaxation time of four distinct samples with variable interparticle center-to-center distances $d_{cc}$ of $\gamma$-Fe$_2$O$_3$ MNPs (diameter $D=4.7\,$ nm) as investigated by Dormann \textit{et al.}~\cite{Dormann1998}: $d_{cc}\approx 21\,$nm (long-range dispersed 'isolated far'
particles IF), $7.3\,$ nm (short-range dispersed 'isolated near' particles IN) and $6.8\,$nm for large conglomerates (clustered 'flocculated' particles Floc). The dry powder sample has the $d_{cc}$ slightly larger than $D$. The parameters derived from the analyses using Eq \ref{model} are enumerated in Table~\ref{tab:fit_parameters}.  The most diluted sample (IF) is the closest to the Néel–Brown ideal behavior. As the interaction strength increases, the relaxation curves develop a slight curvature and the relaxation time grows. From our analysis $q$ values progressively deviate from unity (with $q<1$) as the interaction strength increases and the magnetic correlation increases. When particles are brought closer, dipolar interactions grow and deviations from Boltzmann–Gibbs behavior are observed, as predicted. The higher uncertainty observed in the powder sample could be attributed to exchange-coupling interactions between the particles. The fitted $T_{\mathrm{cut\!-\!off}}$ rises with increasing interaction strength, consistent with a higher temperature for the onset of glassy dynamics in more strongly interacting assemblies. However, the fitted $T_{\mathrm{cut\!-\!off}}$ values lie above the theoretical prediction obtained using the same energy barrier for all samples \(\Delta V/k_B =555\ \mathrm{K}\) (the value found by Dormann \emph{et al.} for the ultra-dilute IF sample), as shown in Fig. ~\ref{fig:Testrela_Tfreezing_sidebyside}. This discrepancy indicates that the effective energy barrier is progressively underestimated when interactions grow. Dormann \emph{et al.} attributed the barrier corrections found for IN and Floc samples using the (DBF) model to changes in the surface anisotropy induced by interactions: as interparticle coupling alters the magnetostatic state of surface spins, the effective barrier experienced by the moments is modified. In short, increasing interaction strength appears to raise the effective energy barrier, which explains why the theoretical curve with similar $q$'s computed with the IF value \(\Delta V/k_B =555\ \mathrm{K}\) underestimates the observed \(T_{\mathrm{cut\!-\!off}}\). The pre-factor $\tau_0$ lies in the expected range for MNPs (typically $\sim 10^{-12}$–$10^{-9}\,$s).

 For comparative purposes, the inset of Fig.~\ref{fig:relaxation-fits-four-samples} shows a fit of the Powder sample using the classical spin-glass critical law,
$\tau = \tau^{*}\left(T_g/(T-T_g)\right)^{z\nu}$,
as previously applied by Dormann \emph{et al.}, yielding $T_g = 130 \pm 4$ K. This characteristic temperature lies within the same range as the $T_{\mathrm{cut\text{-}off}}$ obtained from the Tsallis-based analysis, indicating a consistent temperature scale for the onset of collective freezing effects. Dormann \emph{et al.} also reported independent signatures of collective spin--glass-like freezing near this temperature, including (i) an increase in the nonlinear susceptibility, extracted from the low-field expansion of the MNP magnetization upon cooling toward $T_g \approx 130$ K, and (ii) pronounced waiting-time (aging) dependence in the zero-field-cooled relaxation of the Powder sample at temperatures of 80 K, 90 K, and 118 K, features that are much weaker or absent in the more dilute samples.


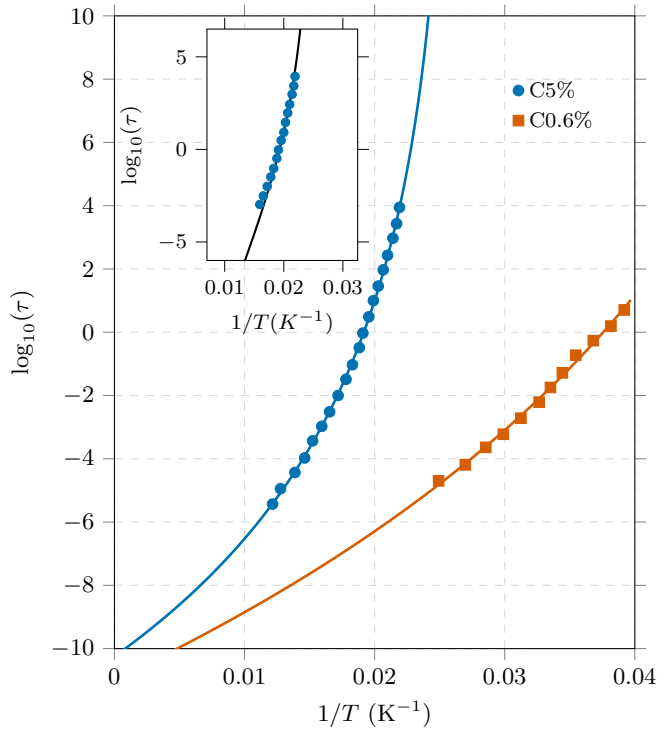
\begin{figure}[h!]
  \centering
  \begin{tikzpicture}

    \begin{axis}[
      name=main,
      width=8.5cm,
      height=10cm,
      xlabel={$1/T\ (\mathrm{K}^{-1})$},
      ylabel={$\log_{10}(\tau)$},
      xmin=0.0, xmax=0.04,
      ymin=-10, ymax=10,
      grid=both,
      grid style={dashed,gray!30},
      tick align=outside,
      scaled x ticks = false,
      tick scale label code/.code={},
      x tick label style={
        /pgf/number format/fixed,
        /pgf/number format/precision=3
      },
      legend style={
        at={(0.75, 0.915)}, 
        anchor=north west, 
        font=\footnotesize,
        draw=none,
        fill=none,
      },
      legend cell align={left},
    ]

      \addplot[only marks, mark=*, color=accessibleBlue,
        mark options={fill=accessibleBlue}]
        table [x index=0,y index=1] {data/AmostraXData.dat};
      \addlegendentry{C5\%} 
      
      \addplot[solid, line width=1pt, color=accessibleBlue, forget plot] 
        table [x index=0,y index=1] {data/AmostraXFit.dat};
      

      \addplot[only marks, mark=square*, color=accessibleRed,
        mark options={fill=accessibleRed}]
        table [x index=0,y index=1] {data/AmostraYData.dat};
      \addlegendentry{C0.6\% } 
      
      \addplot[solid, line width=1pt, color=accessibleRed, forget plot] 
        table [x index=0,y index=1] {data/AmostraYFit.dat};
      

    \end{axis}

    \begin{axis}[
      at={(main.north west)},
      anchor=north west,
      xshift= 35pt, yshift=-5pt,
      width=0.2\textwidth,
      height=0.26\textwidth,
      xlabel={$1/T (K^{-1})$},
      ylabel={$\log_{10}(\tau )$},
      every axis label/.append style={font=\footnotesize},
      scaled x ticks = false,
      scaled y ticks = false,
      xmin=0.007, xmax=0.0325,
      ymin=-6, ymax=6.5,
      tick align=outside,
      axis background/.style={fill=white},
      tick style={black},
      x tick label style={
        /pgf/number format/fixed,
        /pgf/number format/precision=3
      },
      every tick label/.append style={font=\footnotesize}, 
    ]
      \addplot[only marks, mark=*, color=accessibleBlue,
        mark options={fill=accessibleBlue,scale=0.8}]
        table [x index=0,y index=1] {data/Sample1SGData.dat};
      \addplot[solid, line width=0.8pt, color=black]
        table [x index=0,y index=1] {data/Sample1SGFit.dat};
      
        
    \end{axis}

  \end{tikzpicture}
  \caption{Relaxation time versus inverse temperature for the samples C5\% and C0.6\%. 
The main plot shows fits to Eq.~\eqref{model}. 
The inset displays a fit to the spin-glass equation 
for the C5\% sample, yielding the parameters 
$\tau^{*} = 10^{-6\pm1}$ s, $T_g = 40\pm2$ K, and $z\nu = 12\pm3$. 
Data reproduced from Fig.~2 of Djurberg \emph{et~al.} \cite{Djurberg1997}.}
\label{fig:relaxation-with-sg-inset}
\end{figure}

In Figure~\ref{fig:relaxation-with-sg-inset}, results reported by C. Djurberg \emph{et~al.} \cite{Djurberg1997} are presented, with analyses using Eq.~\eqref{model}. Specifically, C. Djurberg \emph{et~al.} examined two different samples of ultrafine Fe--C particles characterized by a median volume \(V=5.3\times10^{-26}\,\mathrm{m}^3\) (diameter \(d\approx4.7\ \mathrm{nm}\)) with varying carbon volume concentrations: one sample at 5\% and the other at 0.6\%. Results obtained from fitting  Eq.~\eqref{model} are summarized in Table~\ref{tab:Hansen_parameters}. As the concentration increases, the entropic index $q$ once more deviates from 1 (decreasing), consistent with increased interaction effects. In the case of the 0.6\% concentration sample, C. Djurberg \emph{et~al.} suggested the potential occurrence of particle agglomeration. This phenomenon might elucidate the observed deviation from ideal superparamagnetic linearity and corresponds with the deduced entropic index $q$, when taking into account the model proposed herein. For comparison, the spin–glass critical law was used to fit the original data and temperature range chosen by the authors in \cite{Djurberg1997}, but the fit is unsatisfactory, as shown in the inset of Fig.~\ref{fig:relaxation-with-sg-inset}. Narrowing the fit to \(45\text{--}55\ \mathrm{K}\) improves agreement, bringing data closer to the glass transition temperature $T_g$. The shorter-range fit yields
$\tau^{*} =10^{-6.2\pm0.3}\,\mathrm{s}$,
$T_{g} = 37\pm2\ \mathrm{K}$,
$z\nu = 15\pm3$.
Regardless of the selected interval, the derived $T_g$ values fall within the same temperature range as the $T_{\mathrm{cut\text{-}off}}$ obtained from fitting the complete 45--85\,K interval using Eq.~\eqref{model}, as summarized in Table~\ref{tab:Hansen_parameters}. Additionally, C.~Djurberg \emph{et~al.} reported magnetic aging in the $C5\%$ sample at $T=40\ \mathrm{K}$ and $T=30\ \mathrm{K}$, whereas no aging was observed for the time window of the experiment at $T=50\ \mathrm{K}$. This finding supports the existence of the superspin-glass behavior for this composition lying in the $30\text{-}40\ \mathrm{K}$ range.

\begin{table}[h!]
\centering
\caption{Parameters obtained from fitting the Eq~\eqref{model}}
\label{tab:Hansen_parameters}
\begin{tabular}{|l|c|c|c|c|}
\hline
\textbf{Sample} & \textbf{$\log_{10}(\tau_0)$} & \textbf{$T_{cut-off}$ [K]} & \textbf{$T^{*}$[K]} & \textbf{$q$} \\
\hline
C5\% & -10.3 $\pm$ 0.2 & 38.4 $\pm$ 0.4 & 2.5 & 0.939 $\pm$ 0.002 \\
\hline
C0.6\%   & -10.9 $\pm$ 0.7 & 15 $\pm$ 2 & 0.8 & 0.96 $\pm$ 0.01 \\
\hline
\end{tabular}
\end{table}

In conclusion, the utilization of Brown's framework to model the magnetic behavior of MNPs with dipolar interactions provides a meaningful contribution to understanding the transition from weakly to strongly interacting magnetic systems, extending to spin-glass transitions.
When interparticle interactions are negligible, the relaxation of magnetic nanoparticles follows the standard Néel–Brown law as derived from escape‐rate theory: a single energy barrier \(\Delta E\) (or a narrow distribution of barriers) leads to a narrow distribution of relaxation times and an Arrhenius‐type plot that yields physically reasonable values for the anisotropy constant and the prefactor \(\tau_0\). As interactions become weak but nonzero, for example in dilute dispersions, the Néel–Brown framework could still represent experimental results with minor modifications: dipolar or exchange coupling shifts the effective barrier up (DBF) or down (MT) without altering the fundamental Boltzmann‐law kinetics, or can be absorbed into a mean‐field correction (VF or \(T^*\) models) that effectively modifies the thermal energy scale \(k_B  T\). Hence, existing theories and models provide a valuable approximation for the behavior of MNPs with weak magnetic interactions,  but they fail to clarify the transition from weak to strong interaction behaviors. The DBF, MT, and VF (or \(T^*\)) models assume thermodynamic equilibrium under the Boltzmann-Gibbs distribution, which holds only for non-interacting systems. This assumption becomes increasingly limited for ensembles of dipole-coupled magnetic nanoparticles, where long-range correlations and collective interactions lead to deviations from Boltzmann statistics. Within this context, the theoretical framework presented here provides an effective description of the experimentally observed relaxation dynamics across regimes ranging from weak to strong interactions, capturing both the decrease and the increase of the relaxation time with growing dipolar coupling.

Moreover, the emergence of a cut-off temperature, $T_{\mathrm{cut\text{-}off}}$, naturally characterizes the onset of glassy freezing dynamics within the Tsallis-based formalism and offers an alternative interpretation of experimental relaxation data, contributing to ongoing discussions on collective freezing phenomena in interacting magnetic nanoparticle systems.

\section{Acknowledgments}
This study was mainly supported and performed under the auspices of the São Paulo Research Foundation (FAPESP) under Grant No. 2024/03819-5 and 24/00998-6. We express our gratitude to Prof. Fernando Fabris.

\section{Data availability}
 The data are available from the corresponding author upon a reasonable request.

\bibliography{bibliografia.bib}

\section{End Matter} \label{End.Matter}

\subsection{Tsallis Statistics}

The entropy of a system, as characterized by the Tsallis distribution, is expressed as:
\begin{equation}\nonumber
S_q = k_B  \frac{1 - \sum_i p_i^q}{q - 1},
\end{equation}
where $q$ denotes the non-extensivity parameter which characterizes the degree of correlation amongst the components of the system. When condition $q \to 1$ is met, the Boltzmann-Gibbs entropy is recovered, thereby ensuring compatibility with traditional statistical mechanics applicable to extensive systems. This non-extensivity is demonstrated through the property specified in 
\begin{equation}\nonumber
\dfrac{S_q(A+B)}{k_B } = \dfrac{S_q(A)}{k_B } + \dfrac{S_q(B)}{k_B } + (1-q) \dfrac{S_q(A)}{k_B } \dfrac{S_q(B)}{k_B },
\end{equation} where $S(A)$ and $S(B)$ represent the entropies of two distinct $\it{independent}$ systems. The maximization of Tsallis entropy under the constraints given by 
\begin{equation}\nonumber
\sum_i^W p(x) = 1 \quad \text{and} \quad \dfrac{\sum_{i=1}^W p_i^q E_i}{\sum_{i=1}^W p_i^q} = U_q,
\end{equation} where $E_i$ is the energy spectrum and $U_q$ is the q-expectation internal energy, resulting in the generalized probability distribution \cite{Tsallis1998ConstraintsNonextensive}: \begin{equation}\label{FP_geral_sol}\nonumber
p_i = \frac{1}{Z'_q} \left[ 1 - (1-q) \beta'_q E_i \right]_{+}^{\frac{1}{1-q}}
\end{equation} 
 Here, $\beta'_q$ is the Lagrange multiplier associated with the energy constraint, related to  $\beta=1/(k_B T)$ via the self-consistent relation:
$$\beta_q'=\dfrac{\beta}{\sum_{i=1}^W p_i^q + (1-q)\beta U_q},$$
With the equilibrium distribution $p_i= \dfrac{e_q^{-\beta'_qE_i}}{Z'_q}$, the partition function can be determined through:
\begin{equation}\label{partição}\nonumber
    Z'_q\equiv \sum_i^W e_q^{-\beta'_qE_i}
\end{equation}
The connection with thermodynamics is established by proving that:
\begin{equation}\nonumber
    \dfrac{1}{T}=\dfrac{\partial S_q}{\partial U_q}
\end{equation}
The condition $[f]_+=max\{f,0\}$ ensures only positive values for the probabilities because sufficiently high energies can produce negative values for the distribution, thus presenting a cut-off condition for the 
q-exponential function $e_q^x$, which can be defined as:
\begin{equation}
    e_q^x = 
\begin{cases} \nonumber
\left(1 + (1-q)x\right)^{1/(1-q)},\text{if } 1 + (1-q)x > 0, \\ 
0, \text{if } q < 1 \text{ and } x < -(1-q)^{-1},  
\end{cases}
\end{equation}
where the fulfillment of condition $1 + (1-q)x > 0$ is necessary to assure the existence of the function. When $ q > 1 $, the domain is constrained to $ x \in \left(-\infty, (q-1)^{-1}\right) $, and the function diverges as $ x \to (q-1)^{-1}$ is approached. Conversely, for the case of $ q < 1 $, the function is defined across all values of $x$, though it nullifies when $ x < -(1-q)^{-1} $. When approaching the limit of $ q \to 1 $, the definition simplifies to the classical exponential form $ e_q^x \to e^x $, ensuring consistency with traditional theory.

\subsection{Analyses Methods }

The fits were performed using non-linear least squares. For each dataset \(T^{*}\) was held fixed, chosen from theoretically meaningful values, while the remaining free parameters were determined by the fit. It has been shown in Fig \ref{fig:Testrela_Tfreezing_sidebyside} that both $T^{*}$ and \(T_{\mathrm{cut\!-\!off}}\) vary with temperature, with a more pronounced variation the further the \(q\) parameter is from \(1\). Nevertheless, the corresponding range of blocking temperatures remains 
limited. As a result, treating $T_{\mathrm{cut\!-\!off}}$ as an effective 
constant still provides a reliable estimate of the collective freezing 
temperature when compared with other approaches, particularly for strongly 
interacting systems close to the crossover regime. When the fitted temperature 
window lies far from the freezing temperature, an underestimation of $T_g$ is 
expected, as illustrated for values of $q$ close to unity in 
Fig.~\ref{fig:Testrela_Tfreezing_sidebyside}. Within these bounds, however, the 
model remains internally consistent and physically meaningful.

\end{document}